# Photonic Time Crystals and Parametric Amplification: similarity and distinction

*Jacob B Khurgin*

Department of Electrical and Computer Engineering, Johns Hopkins University, Baltimore, Maryland 21218, USA  E-mail jakek@jhu.edu

**Abstract:** Photonic Time crystals (PTC) arise in time-modulated media. PTCs are manifested by the generation and amplification of so-called "time reversed" waves propagating in the direction opposite to the incoming light. Superficially, the observed phenomenon bears resemblance to the widely known phenomena of optical parametric generation (OPG) and amplification (OPA) using second or third order optical nonlinearities. I show that while indeed the same physical mechanism underpins both PTC and OPA , the difference arises from the boundary conditions. Thus , while dispersion for both PTC and OPA exhibit the same bandgap in momentum space, only in the case of PTC can one have propagation in that bandgap with exponential amplification. I also show that PTC can be engineered with both second and third order nonlinearities, and that rather unexpectedly, modulating permittivity on the ultrafast (few fs) rate is not a necessity,  and that one can emulate all the  PTC  features using materials with a few picoseconds response time commensurate with the propagation time through the medium.

## 1. Introduction

Diverse phenomena associated with light propagation in materials with temporary modulated optical properties have moved into the focus of attention of the photonic community in recent years[1-5]. One can argue that exploring the fourth, temporal dimension is a natural extension of the developments in the research on three and two dimensional spatially modulated metamaterials and photonic crystals. Alternatively, one can say that photonics offers an ideal testing platform for exploring temporal modulation effects previously studied theoretically, such as formation of Floquet crystals.[6] Last, but not least, it is in the optical domain that time modulated phenomena may segue into practical applications from optical isolators [7, 8] to generating entangled photons.  At this time, time modulated phenomena such as time reflection/refraction that is a precursor to the  subject of this essay  photonic  time crystals (PTC) [5, 9, 10] have been demonstrated in the microwave domain [11], while progress in the optical domain has been so far impeded by the difficulty of achieving fast (on the scale of an optical period, i.e. a few femtoseconds) modulation of permittivity in available materials, despite significant progress achieved in relatively new materials, transparent conductive oxides[5].  Yet  with optimistic but  not entirely unreasonable expectation that sooner or later time reflection and time crystals in optical domain will become feasible[5, 12], it would be logical to attempt to ascertain what, if any, benefits such developments will bring beyond what had been achieved and utilized routinely.

Since the most (and perhaps the only) realistic way of achieving modulation of the permittivity on the femtosecond scale is by doing it with light, the time modulation phenomena must be closely related to the widely explored nonlinear optical phenomena based on second $\chi^{(2)}$ and third $\chi^{(3)}$ nonlinear susceptibilities[13]. These phenomena are optical parametric generation (OPG) and amplification (OPA), four wave mixing (FWM) and phase conjugation (PC). The similarities have been duly noted and questions are often posed, asking what is the difference between parametric processes and time modulated phenomena, especially PTC? In each of these phenomena the new (idler) wave is generated while the energy of the strong pump wave is transferred to that idler as well as to the original signal wave, so that both of them experience amplification. Some answers to those who doubt the novelty of PTC have been provided[10], but these answers have not been complete and quantitative, as, for instance, no fair comparison of either efficiency or bandwidth of PTC vs. that of OPA has been made in the literature. Similarly, since formation of PTC using $\chi^{(3)}$ involves four photons, no detailed analysis of how it relates to FWM and PC[14] has been presented. Lastly it is often pointed out [10] that unlike more conventional parametric processes, phase matching plays no role in formation of PTC, which, with all due respect, is not correct as the lack of phase matching in time rather than in space does negatively affect the efficiency of amplification in PTC.

In this exercise I attempt to clarify the issues outlined above and provide an answer to the question of what unites and what separates PTC and parametric processes by showing (using only elementary analytical derivations) that the main difference between the two is only in the boundary conditions. While energy is conserved in OPA/OPG, in PTC it is only momentum that stays unperturbed. The bandgap in momentum space arises in both OPA and PC, but energy conservation at the boundary prevents operation inside the bandgap. I also show that, when third order nonlinearity is used, many PTC features (bandgap and amplification) can be observed without ultrafast modulation of permittivity, by relying on transient grating oscillating at the beat frequency between pump and signal.

## 2. Backward parametric generation and amplification

To commence, one may consider the second order nonlinear parametric processes. Quite a few different geometries can be used to observe parametric phenomena, but to make a fair comparison with PTC, it would be best to focus on the backward OPA and OPG in which the signal and idler waves are counterpropagating and are phase conjugated. Backward OPA was first proposed by Harris as early as 1966[15] and later modified by Ding et al [16] for the case of transverse pumping and waveguide propagation in which the phase matching requirements are relaxed and the propagation distance is sufficient to attain meaningful amplification. OPG in this scheme was experimentally demonstrated in 2006[17]. This geometry can be adapted to operating in either the OPA or the PTC regime, and I chose it to perform a comparative analysis of the two. While the theory of backward OPA is straightforward and well-explored, I find it vital to provide a concise derivation in order to pinpoint the key distinction between OPA and PTC. This distinction is in the opposite signs of the temporal and spatial derivatives in coupled equations that leads to the oscillatory character of propagation in backward OPA vs. exponential growth in PTC.

As shown in Fig.1a, the waveguide made of material with nonzero second order susceptibility $\chi^{(2)}$ supports a single mode at frequencies $\omega_1$ (signal) and $\omega_2$ (idler) propagating in opposite directions

$$\boldsymbol{E}_{1,2} = A_{1,2}\boldsymbol{f}(x,y)e^{i(\pm k_{1,2}z - \omega_{1,2}t)} + A_{1,2}^{*}\boldsymbol{f}(x,y)e^{i(\mp k_{1,2}z + \omega_{1,2}t)} \qquad (1)$$

where $A_{1,2}$ are the amplitudes, vector $\boldsymbol{f}(x,y)$ is the normalized mode profile; propagation constants are $k_{1,2} = \omega_{1,2}n_{eff}/c$, and $n_{eff}$ is the effective index. The pump wave propagates in the direction $x$ normal to $z$, $\boldsymbol{E}_p = A_3 e^{i(k_3 x - \omega_3 t)}\hat{\boldsymbol{y}} + A_3^{*}e^{-i(k_3 x - \omega_3 t)}\hat{\boldsymbol{y}}$ where $\hat{\boldsymbol{y}}$ is a unit vector of pump wave polarization. Without loss of generality one can take $A_3$ to be real. As a result of parametric interaction with signal nonlinear polarization at the difference frequency, $\omega_2 = \omega_3 - \omega_1$ is generated,

$$\boldsymbol{P}_{NL}(\omega_2) = \varepsilon_0\tilde{d}\boldsymbol{E}_P\boldsymbol{E}_1 = \varepsilon_0\tilde{d}A_3 A_1^{*}e^{i(-k_1 z - \omega_2 t)}\hat{\boldsymbol{y}}\boldsymbol{f}(x,y)e^{-ik_3 x} + c.c., \qquad (2)$$

where $\tilde{d}$ is the tensor of second order susceptibility. A crucial point here is that we consider the case when the time modulation is continuous in time (i.e. starting long before the signal wave first arrives at the boundary at z=0) but restricted in space to the extent of the waveguide between z=0 and z=L. Then energy conservation dictates that nonlinear polarization engender the backward propagating wave at the same frequency $\omega_2$.

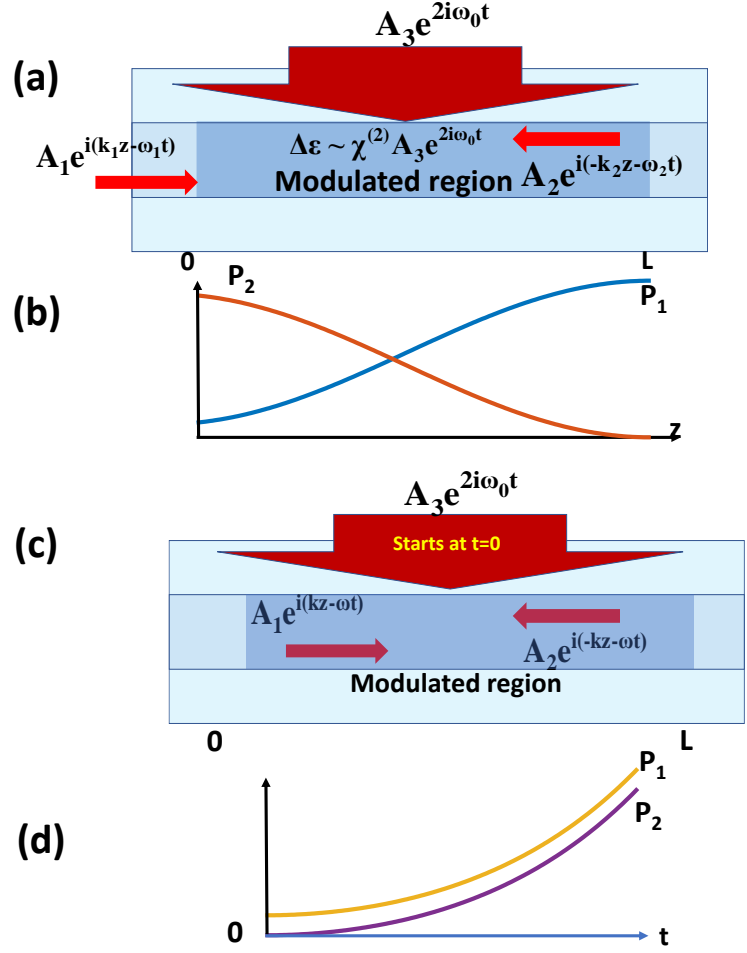

**Figure 1.** (a) Nonlinear waveguide for observation of OPA  The signal enters the waveguide after permittivity variation at frequency $2\omega_0$ has been established by pump $A_3$. (b) variations of signal ($P_1$) and idler ($P_2$) along the length of waveguide. (c) Arrangement required for observation of PTC. Signal wave is already inside the waveguide by the time permittivity modulation commences. (d) Exponential growth of signal and idler waves in time.

The nonlinear interaction between the signal wave and pump gives rise  to nonlinear polarization at the signal frequency,

$$\boldsymbol{P}_{NL}(\omega_1) = \varepsilon_0 \tilde{d} \boldsymbol{E}_p \boldsymbol{E}_2 = \varepsilon_0 \tilde{d} A_3 A_2^* e^{i(-k_2 z - \theta t)} \hat{\boldsymbol{y}} f(x,y) e^{ik_3 x} + c.c. \tag{3}$$

One can now substitute nonlinear polarizations (2) and (3) into the wave equation

$$\nabla^2 \boldsymbol{E} - \frac{n^2}{c^2} \frac{d^2 \boldsymbol{E}}{dt^2} = \frac{1}{c^2 \varepsilon_0} \frac{d^2 \boldsymbol{P}_{NL}}{dt^2} \tag{4}$$

The boundary conditions indicate that amplitudes $A_{1,2}$ are only the functions of coordinate $z$. Then, using a slow variable approach (neglecting second derivatives) used in every textbook, multiplying both sides by $f(x,y)$ and performing integration over transverse coordinates x and y, we obtain the usual set of coupled equations

$$\frac{dA_1}{dz} = i\frac{\omega_1 d_{eff} A_3}{2n_{eff}c} A_2^* e^{i2\Delta kz}$$

$$\frac{dA_2^*}{dz} = i\frac{\omega_2 d_{eff}^* A_3}{2n_{eff}c} A_1 e^{-i2\Delta kz}$$

(5)

where $d_{eff} = \iint f(x,y)\tilde{d}\tilde{y}f(x,y)e^{ik_3 x}dxdy = |d_{eff}|e^{i\varphi}$ is the effective second order susceptibility, $n_{eff}^2 = \iint n^2(x,y)f^2(x,y)e^{ik_3 x}dxdy$ is the square of effective mode index, momentum mismatch is

$$2\Delta k = k_2 - k_1 = 2n_{eff}\Delta\omega/c,$$

(6)

and $\Delta\omega = \omega_0 - \omega_1$, where $\omega_0 = \omega_p/2$ is the frequency of degenerate OPA. Normalizing amplitudes to the photon numbers and eliminating the imaginary numbers by introducing new variables $a_1 = A_1 e^{i\varphi}\omega^{-1/2}$ and $a_2 = iA_2^*\omega_2^{-1/2}$, one arrives at the well-known set of coupled equations

$$\frac{da_1}{dz} = \kappa a_2 e^{i2\Delta kz}$$

$$\frac{da_2}{dz} = -\kappa a_1 e^{-i2\Delta kz}$$

(7)

where the coupling coefficient is $\kappa = (\omega_1\omega_2)^{1/2}|d_{eff}|A_3/2n_{eff}c$ .

Before proceeding further, it is worthwhile to note the opposite signs of derivatives in the first and second equations in (7) – clearly this the natural consequence of the opposite directions in space propagation of signal and idler waves. Consequently, the solution is harmonic and not exponential. I now proceed by introducing

$$a_1 = b_1 e^{i\Delta kz}, \ a_2 = b_2 e^{-i\Delta kz},$$

(8)

and substituting it into (7) end up with

$$\frac{db_1}{dz} + i\Delta k b_1 = \kappa b_2$$

$$\frac{db_2}{dz} - i\Delta k b_2 = -\kappa b_1$$

$(9)$

Substituting the expected solution $b_{1,2} \sim e^{j\beta z}$ yields characteristic equation $\beta^2 = \kappa^2 + \Delta k^2$ indicating that the propagation constant is always real. Applying the boundary conditions $b_1(0) = A_0; b_2(L) = 0$ I obtain

$$b_1 = A_0 \cos \beta z + \frac{\tan \beta L - i\dfrac{\Delta k}{\beta}}{1 + i\dfrac{\Delta k}{\beta}\tan \beta L} A_0 \sin \beta z$$

$$b_2 = A_0 \kappa^{-1}\left[ -\beta + i\Delta k \frac{\tan \beta L - i\dfrac{\Delta k}{\beta}}{1 + i\dfrac{\Delta k}{\beta}\tan \beta L} \right] \sin \beta z + A_0 \kappa^{-1}\left[ \beta \frac{\tan \beta L - i\dfrac{\Delta k}{\beta}}{1 + i\dfrac{\Delta k}{\beta}\tan \beta L} + i\Delta k \right] \cos \beta z,$$

$(10)$

indicating the oscillatory character of the amplitudes as shown in Fig. 1b.

Let us now obtain the expression for the dispersion of the wavevector inside the nonlinear medium, $k_{in} = k_1 + \Delta k \pm \beta$ using (6) and (10)

$$ck_{1,2in} / n_{eff} = \omega_0 \mp \sqrt{\kappa_t^2 + (\omega_0 - \omega_{1,2})^2},$$

$(11)$

where the temporal coupling coefficient is $\kappa_t = c\kappa / n_{eff} = (\omega_1 \omega_2)^{1/2} |F| d_{eff} A_3 / 2n_{eff}^2$. The dispersion is plotted in Fig.2 and the bandgap around $k_0 = \omega_0 n_{eff} / c$ whose width is $2\kappa$ can be clearly spotted. Since the boundary conditions allow change of momentum, the light whose wavevector outside the time modulated region $k_1 = \omega_1 n_{eff} / c$ falls into the bandgap changes its momentum to $k_{1,in}$ (11) when it enters the medium and then changes it back upon the exit. The situation is quite different in the case of PTC.

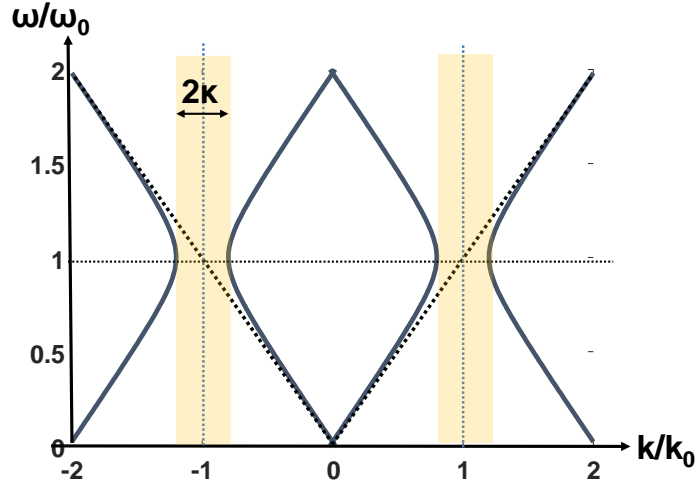

**Figure 2** Dispersion of light inside the time modulated medium. For the OPA arrangement, the boundary conditions prevent light from being inside the bandgap, but for PTC it is allowed.

### 3. Time crystals in second order nonlinear materials

A time crystal is formed when time modulation occurs after the signal has already entered the nonlinear region and the modulation has commenced afterwards, as shown in Fig.1c. In this situation the wavevector must be conserved and the forward wave $E_1 = A_1 e^{i(kz - \omega t)} + c.c.$ can only be efficiently coupled into the counter-propagating wave $E_2 = A_2^* e^{i(kz + \omega t)} + c.c.$ when the pump wave at frequency $\omega_3 = 2\omega_0$ is turned on. Once the pump is on, translational symmetry assures that amplitudes of the waves are the functions of time only. In place of equations (2,3) we now have the following expressions for nonlinear polarizations propagating forward (1) and backward (2)

$$\boldsymbol{P}_{NL2(1)} = \varepsilon_0 \tilde{d} \boldsymbol{E}_p \boldsymbol{E}_{1(2)} = \varepsilon_0 \tilde{d} A_3 A_{1(2)}^* e^{i(\mp kz + \omega t - \omega_p t)} \hat{\boldsymbol{y}} \boldsymbol{f}(x,y) e^{-ik_3 x} + c.c. \tag{12}$$

where the phase mismatch $\Delta\omega = \omega_0 - \omega$ now occurs in time rather than in space.

It should be noted that (12) can be easily generalized for the case of arbitrary modulation of the dielectric constant $\varepsilon(t) = \varepsilon_0 [\overline{\varepsilon}_r + \Delta\varepsilon_r(t)]$ as long as time dependent tensor of dielectric constant is harmonic $\Delta\varepsilon_r(t) = \Delta\varepsilon_{r,\max}(x,y)\cos\omega_p t$. The polarization induced by the permittivity modulation (subscript "*tve*" standing for "time variable epsilon") then becomes simply

$$\boldsymbol{P}_{tve2(1)} = \frac{1}{2} \varepsilon_0 \Delta\varepsilon_{r,\max} A_{1(2)}^* e^{i(\mp kz - \omega t - 2\Delta\omega t)} \boldsymbol{f}(x,y) + c.c. \tag{13}$$

and all the ensuing derivations are applicable to the most general case as long as second rank tensor $\Delta\varepsilon_{r,\max}/2$ is used in place of second rank tensor $\tilde{d}A_3\hat{y}$. In fact, $\tilde{d}E_p$ is precisely the time dependent change of **linear** susceptibility (or **linear** dielectric constant) caused by the optical field $\varepsilon_r(t) \sim \overline{\varepsilon}_r + \tilde{d}E_p(t)$. In our case pump wave is much stronger than signal and idler and we neglect the terms such as second harmonic and sum frequency generation, therefore from the point of view of signal and idler the system is linear.

In case of Pockels effect one would have $\Delta\varepsilon_{r,\max} \sim \overline{\varepsilon}_r^2 rE$ where $r$ is Pockels coefficient and $E$ is the low frequency electric field, for third order nonlinearity it would be $\Delta\varepsilon_{r,\max} \sim 2nn_2I$, where $n_2$ is nonlinear index and $I$ is intensity, and for acousto-optic effect $\Delta\varepsilon_{r,\max} \sim \overline{\varepsilon}_r^2 PS$, where $S$ is the strain and $P$ is a photoelastic coefficient. Of all the methods of achieving time crystals in optical range, which requires permittivity modulation at the optcal frequencies, second order nonlinearity is probably the most realistic and this is why it is the focus of this study.

We proceed now substituting (13) into the wave equation (4), then using the fact that the amplitudes of counterpropagating waves depend only on time and not space and assuming that amplitude changes are small over optical period (neglecting second order derivatives in time $d^2A/dt^2 \ll \omega dA/dt$ ),we obtain

$$\nabla^2 \boldsymbol{E}_{1(2)} - \frac{n^2}{c^2}\frac{d^2\boldsymbol{E}_{1,2}}{dt^2} = \left[\left(\nabla_{xy}^2 - k^2 + \frac{n^2}{c^2}\omega^2\right)A_{1(2)} + \frac{n^2}{c^2}2i\omega\frac{dA_{1(2)}}{dt}\right]f(x,y)e^{i(\pm kz - \omega t)} + c.c. =$$
$$= -\frac{\omega^2}{c^2}\tilde{d}A_3 A_{2(1)}^* e^{i(\pm kz - \omega t - 2\Delta\omega t)}f(x,y)\hat{y}e^{-ik_3 x} \tag{14}$$

The term in the parenthesis on the left-hand side is equal to zero as a solution of unperturbed wave equation in the waveguide. Then cancelling a number of terms on both sides , integrating, and introducing effective index $n_{eff}$ and second order susceptibility $d_{eff}$ as before, and finally taking complex conjugate of one of the coupled equations we obtain

$$\frac{dA_1}{dt} = i\frac{\omega\,d_{eff}A_3}{2n_{eff}^2}A_2^* e^{-i2\Delta\omega t}$$
$$\frac{dA_2^*}{dt} = -i\frac{\omega\,d_{eff}^* A_3^*}{2n_{eff}^2}A_1 e^{i2\Delta\omega t} \tag{15}$$

If one compares (15) with (5), one can see that the main difference is the change of sign in the second of the coupled equations. This is easily explained ,in space, two coupled waves propagate in the opposite directions, while in time they obviously move in the same direction (one should not take the term "time reversed" literally), hence the time derivatives have the same sign. Following the derivations of the previous section , i.e. introducing $b_1 = A_1 \omega^{-1/2} e^{i\varphi - i\Delta\omega t}$ and $b_1 = iA_2^* \omega^{-1/2} e^{i\Delta\omega t}$ we obtain a new set of coupled equations

$$\frac{db_1}{dt} - i\Delta\omega b_1 = \kappa_t b_2$$

$$\frac{db_2}{dz} + i\Delta\omega b_2 = \kappa_t b_1$$

(16)

Where $\kappa_t = c\kappa / n_{eff} = \omega | d_{eff} | A_3 / 2n_{eff}^2$ It is easy to see that, due to r.h.s. of both equations having the same sign, an exponentially growing solution becomes possible. Substituting $b_{1,2} = b_{1,2} e^{\gamma t}$ into (16) I obtain the characteristic equation $\gamma^2 = \kappa_t^2 - \Delta\omega^2$ . Therefore, as long as $\Delta\omega < \kappa_t$ one has solution that is a sum of exponentially increasing and decreasing waves with exponentially increasing one dominating. With initial conditions $b_1(0) = A_0$ and $b_2(0) = A_0$ it follows that

$$b_1(t) = A_0 \left[ \cosh\gamma t + i\frac{\Delta\omega}{\gamma}\sinh\gamma t \right]$$

$$b_2(t) = A_0 \frac{\kappa}{\gamma}\sinh\gamma t$$

(17)

The complex frequencies of two waves inside the gap can then be found as $\tilde{\omega}_{1,2,in} = \omega_0 \pm j\gamma$ .

When the wave is outside the bandgap, then $\gamma = i\beta_t = i\sqrt{\Delta\omega^2 - \kappa_t^2}$ and the solution is oscillatory,

$$b_1(t) = A_0 \left[ \cos\beta_t t + i\frac{\Delta\omega}{\beta_t}\sin\beta t \right]$$

$$b_2(t) = A_0 \frac{\kappa_t}{\beta_t}\sin\beta_t t$$

(18)

The fields of two counterpropagating waves then become

$$E_1(t) = \frac{A_0}{2}\left[1 + \frac{\Delta\omega}{\beta}\right]e^{i(kz - \omega_0 t + \beta t)} + \frac{A_0}{2}\left[1 - \frac{\Delta\omega}{\beta}\right]e^{i(kz - \omega_0 t - \beta t)} + c.c.$$

$$E_2(t) = \frac{A_0}{2}\frac{\kappa}{\beta}\left[e^{i(-kz + \omega_0 t - \beta t)} - e^{i(-kz - \omega_0 t + \beta t)}\right] + c.c.$$

(19)

That means that, once inside the time modulated interval, the eigenmodes are superpositions of two waves with eigenfrequencies $\omega_{1,2,in} = \omega_0 \pm \sqrt{(\omega - \omega_0)^2 - \kappa_t^2}$ , which are real outside of the bandgap and complex inside of it. Note the important fact that, during the time interval when modulation takes place, one always has $\omega_1 + \omega_2 = 2\omega_0 = \omega_3$ , indicating that energy is conserved as it is transferred from a pump photon to signal and idler photons – a standard condition of all parametric processes. It is only at temporal boundaries when frequencies are no longer conserved – for instance after the pump signal stops, both signal and idler revert to original frequency $\omega$ . Whether one is inside (17) or outside (18) the bandgap , the relation $|b_1|^2 - |b_2|^2 = A_0^2$ indicates that signal and idler photons are generated simultaneously and thus entangled although it can also be interpreted as Minkowski momentum conservation[18, 19]. Also note that, in the vicinity of $\omega = \omega_0 \pm \kappa_t$ , exceptional points[20] occur, with plenty of well-publicized features of dubious practicality[21-23].

Let us now find the dispersion outside the bandgap. Since $k = \omega n_{eff} / c$ , I obtain

$$(\omega_{1,2,in} - \omega_0)^2 + \kappa_t^2 = (kc / n_{eff} - \omega_0)^2, \qquad (20)$$

and

$$kc / n_{eff} = \omega_0 \pm \sqrt{\kappa_t^2 + (\omega_0 - \omega_{1,2,in})^2} \qquad (21)$$

which is of course *identical* to the expression (11) for the dispersion in OPA as shown in Fig.2

## 4. Significance of boundary conditions

To summarize what has been shown so far, the dispersion for time modulated material looks the same whether one considers PTC or backward OPA and energy conservation is maintained as pump photons split into signal and idler with energy conserved. So, why, then, can PTC show exponential growth, while backward OPA cannot? To discern it, consider the difference in boundary/initial conditions. In Fig. 3 a-c I show the development of amplification in OPA. In Fig. 3a the signal light is outside the modulated region and has frequency $\omega_1$ and wavevector $k_1$ is propagating, as expected, situated on the straight line describing unmodulated waveguide dispersion. Now, when the signal crosses the spatial boundary of the modulation region (shown in darker shade), the boundary condition calls for energy, i.e. frequency conservation. (The point $\omega, k$ can only move along horizontal line.) Hence, the momentum becomes $k_{1,in}$ as shown in Fig.3b. If the value of $k_1$ corresponded to the bandgap in the momentum space (as in the case shown in Fig.3), it is now "pushed" outside of it. Inside the modulated region the idler wave with momentum $-k_{2,in}$ is then generated. Both the signal and idler are amplified, and, once they reach their respective boundaries (at z=L and z=0) respectively, the wavevectors revert to their original values $k_1$ and $k_2$ as shown in Fig.3c.

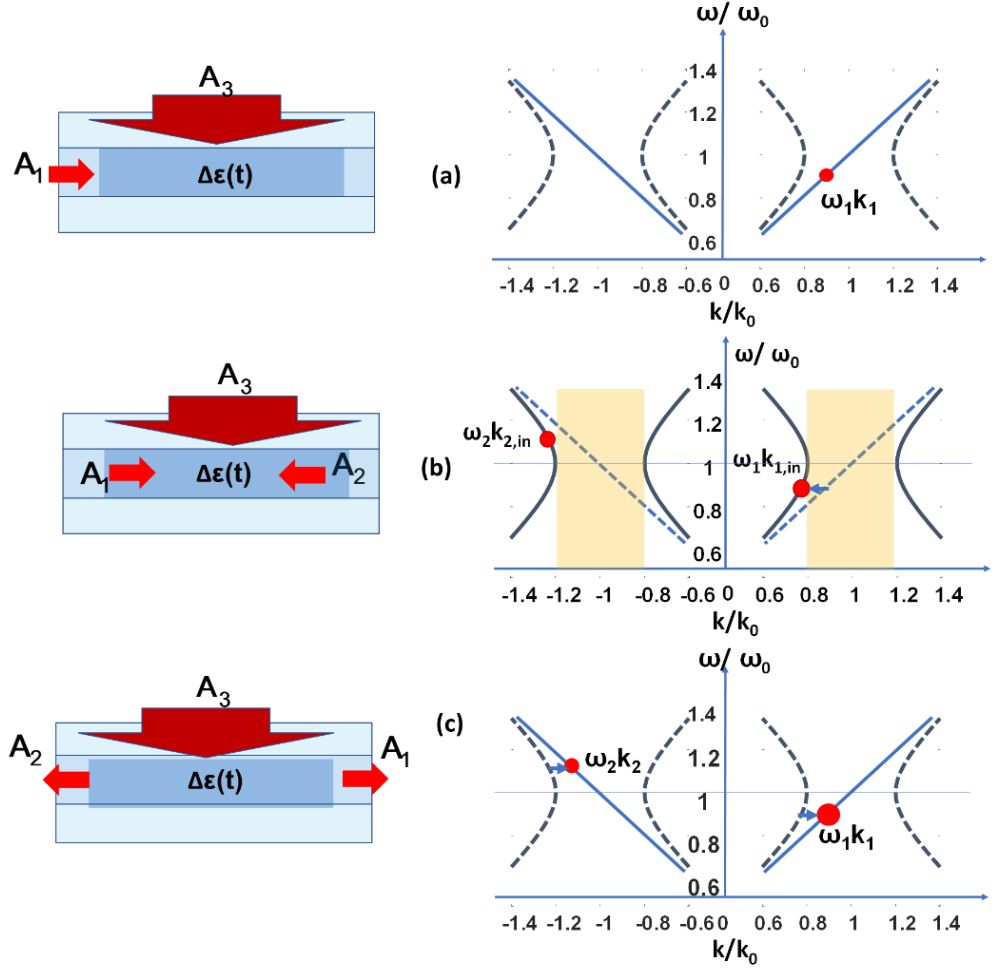

**Figure 3.** Light propagation through the time modulated medium in the OPA arrangement. (a) signal wave is outside the dark shaded modulated region. (b) signal is inside the modulation region and its wavevector changes while its frequency is conserved. A counterpropagating idler is generated. (c) Signal and idler leave the modulated region and their wavevectors return to their values in the unmodulated waveguide.

In case of PTC, light of frequency $\omega$ and momentum $k$ is already inside the active region prior to the pump (Fig.4a), so at time $t=0$, when the modulation commences, the wavevector is preserved and the frequency becomes $\omega_{1,2,in}$ (one can move only in the vertical direction in the dispersion diagram), as shown in Fig.4b. If $k$ happens to be within the bandgap, the frequency is "pushed" inside the gap. The frequencies of the signal and idler are complex, with the same real part $\omega_0$ and imaginary parts describing exponential growth and decay. Once the temporal modulation is complete (t=T) the frequency reverts back to the original $\omega$ as shown in Fig.4c . In other words, the conjugated wave is at the same frequency as incoming wave . <mark>How exactly the frequency shift takes place at the boundary depends on the rate at which modulation field rises and decays, and this aspect merits further in-depth investigation.</mark>

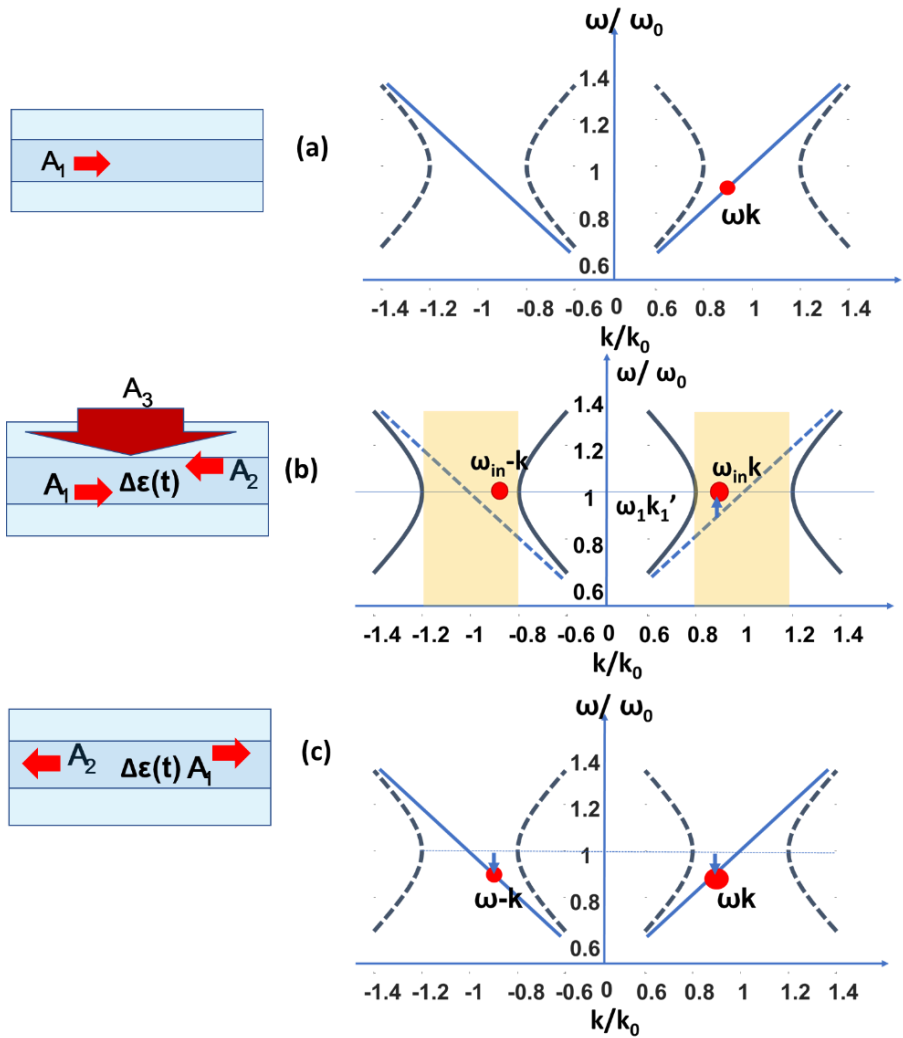

**Figure 4.** Light propagation through the time modulated medium in the PTC arrangement.(a) the signal wave is inside the time modulated region, but the modulation has not commenced yet. (b) Time modulation has commenced and the signal frequency changes to a complex value while its wave vector

is conserved. The counterpropagating idler is generated. (c) Time modulation has stopped, and frequencies of signal and idler revert to the original frequency $\omega$.

Note that, even though energy conservation may not be maintained at each boundary, it is maintained for the entire time T from start to end as one can easily see that $|b_1(T)|^2 - |b_2(T)|^2 = |b_1(0)|^2$ indicating that forward and backward propagating photons are generated at the same rate, as in any parametric process.

## 5. Performance comparison

Having established the similarities and differences between OPA and PTC on the fundamental level, now is the time to compare them on a more practical and application-relevant level, starting from a simple comparison of amplification attainable in either of the schemes. That can be done by considering the fact that, to ensure translational symmetry and momentum conservation, the entire modulation time in OTC cannot exceed the propagation time. Therefore using $L = Tc / n_{eff}$ one can easily see that $\kappa L = \kappa_t T$ and one can obtain from (10)

$$b_{1,out} = b_1(L) = A_0 \left[ \cos \beta_t T + \frac{\tan \beta_t T - i \dfrac{\Delta \omega}{\beta_t}}{1 + i \dfrac{\Delta \omega}{\beta_t} \tan \beta_t T} \sin \beta_t T \right]$$

$$b_{2,out} = b_2(0) = \kappa_t^{-1} A_0 \left[ \beta \frac{\tan \beta_t T - i \dfrac{\Delta \omega}{\beta_t}}{1 + i \dfrac{\Delta \omega}{\beta_t} \tan \beta_t T} + i \Delta \omega \right]$$

(22)

where $\beta_t = \sqrt{\kappa_t^2 + \Delta \omega^2}$. Let us now plot the values of output powers $P_{1,2} = |b_{1,2,out}|^2$ for OPA according to (22) (dashed lines) and for PTC according to (17) (solid lines )for different values of peak parametric single pass gain $\kappa L = \kappa_t T$.

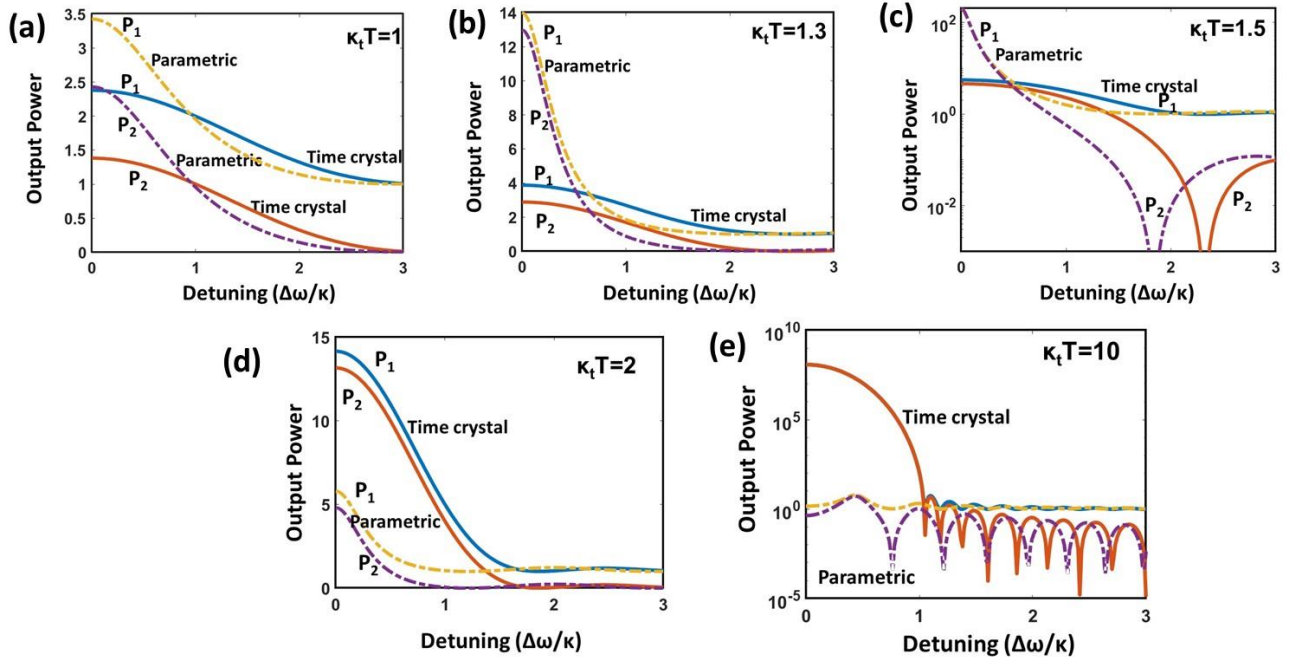

**Figure 5.** Output power relative to input power for OPA (dashed lines) and PTC (solid lines) for different values of a single pass parametric gain $\kappa L = \kappa_t T$.

Prior to doing it, let us ascertain what values of gain are achievable. Taking the effective second order susceptibility on the scale of $d_{eff} |F| \sim 100 \, pm/V$ for GaAs[24] , $L = 1mm$ and pump intensity of 1GW/cm$^2$ ( $A_3 = 5 \times 10^7 V/m$ ) one can obtain $\kappa L \sim 4$ and the width of bandgap $2\kappa_t = \omega |F| \, d_{eff} A_3 / 2\pi n_{eff}^2 \sim 130 GHz$ - this value is less than 0.1% of signal frequency but sufficient to accommodate the entire spectrum of the signal pulse whose duration is less than transit time $T \sim 10 \, ps$. These numbers, as well as experimental results [17], appear to confirm that, with a bit of luck and ingenuity, both OPA and PTC are attainable using a second order nonlinearity. I am not going to further comment on practical issues (such as, for instance focusing the pump onto a stripe with high aspect ratio) , as the goal here is just to provide the comparison of OPA and PTC.

Now, we turn our attention to Fig.5.a where the output intensities of the forward and backward waves are plotted for the case of single pass parametric gain $\kappa_t T = 1$. For these moderate values of gain, the overall output power gain curves exhibit similar smooth dependences on detuning from the central frequency $\omega_0$ . As the parametric gain increases to $\kappa_t T = 1.3$ (Fig.5b) the overall gain for OPA increases near $\omega_0$ . The reason is the presence of positive feedback in space. Obviously, this feedback is unattainable in the time domain as, despite the term "time reversal," no light can travel back in time. Ultimately, when one approaches the threshold condition $\kappa_s T = \pi/2$ (as shown in Fig.5c), optical parametric oscillation commences, at which point the parametric gain gets clamped at threshold value due to depletion of the pump. If the pulse pump is short, however (less than a few round trip pass times), then parametric oscillation will not have time to develop, and one can in principle continue to increase the parametric gain to the value $\kappa_s T = 2$, at which point PTC does show large gain within the bandgap and relatively little gain outside of it (Fig.5d). Finally, assuming that a huge gain of $\kappa_s T = 10$ can be achieved without gain depletion due to amplified spontaneous parametric down conversion, the difference between PTC and OPA becomes obvious – in PTC the overall gain exceeds 20dB inside the bandgap and goes to essentially zero outside of it. Experience shows that overall gain beyond 30dB cannot be achieved due to stimulated emission (optical parametric generation in this case) that depletes the gain. Hence the picture presented in Fig.5e is of mostly academic interest, not to say that it does not have any value, as it elucidates when and where the difference between OPA and PTC becomes perceptible.

From the practical point of view , it has been purported that one can achieve parametric amplification in PTC over a broader bandwidth than in OPA [25] and that parametric gain is frequency-independent as long as it lies within the gap[10].   This is obviously not entirely correct, as gain in PTC declines severely with detuning from center frequency $\omega_0$. Furthermore, in  OPA geometry, even outside  the bandgap the amplification bandwidth is comparable, and gain is larger than in PTC for realistic  values of gain in Fig.5 a-c. If one really wants to achieve a broad bandwidth with OPA, then one should consider  a standard forward OPA operating around degeneracy, where bandwidth is determined by group velocity  dispersion (GVD) $\beta_2$, $\Delta\omega = 2\sqrt{\kappa/\beta_2}$ and amplification bandwidths as broad as 250THz are routinely attainable [26] so that white light can be generated in them [27]. Nothing like that can be obtained in the scheme with counterpropagating signal and idler, in either OPA or PTC arrangement, although counterpropagating geometry may work for wider angular range.

## 6. Time crystal formation  using nonlinear index and its relationship with optical phase conjugation.

Just as using the second order nonlinearity, one may consider using the third order processes[13, 28]. The advantage of the third-order processes is that the phase matching in them is less critical. However, the significant shortcoming is that the strength of the third-order processes, at least in the case of ultrafast ones, is less than that of the second-order processes. But does one really need an ultrafast process to observe the usual manifestations of PTC? Here, I provide a somewhat surprising answer

The third order nonlinear polarization in the presence of a pump wave of frequency $\omega_0$ can formally be written as

$$P_{NL} = \varepsilon_0 3\chi^{(3)} E_P E_P (E_1 + E_2) = \varepsilon_0 3\chi^{(3)} A_3^2 \left( A_1 e^{i(kz+[2\omega_0-\omega]t)} + A_2^* e^{j(kz-[2\omega_0-\omega]t)} \right) e^{-2ik_3x} + c.c., \qquad (23)$$

where only the terms relevant to the formation of PTC (or to PC) have been kept. The factor of 3 is here because one has three distinct sequences in which three waves interact, $E_{1,2}E_P E_P$, $E_P E_{1,2} E_P$, and $E_P E_P E_{1,2}$. These three terms, however, describe different processes and have different values of nonlinear susceptibilities[29]. The first term,

$$P_{NL}^{fast} = \varepsilon_0 \chi^{(3)}(2\omega_0 - \omega, -\omega, \omega_0, \omega_0) A_1 e^{jkz-j\omega t} A_3^* e^{j\omega_0 t} A_3^* e^{j\omega_0 t} e^{2ik_3x} \qquad (24)$$

describes the process in which permittivity is modulated at frequency $2\omega_0$ which in turn modulates the signal wave and scatters it into the counterpropagating idler (and vice versa). The sequence of events is shown in Fig.6a schematically. This process is non resonant and hence instant and relatively weak with $\chi_{fast}^{(3)} \sim 10^{-19} - 10^{-20} m^2/V^2$[30]. Following derivations in the previous section, one can easily obtain the same coupled set of equations (16) with coupling coefficient $\kappa_t = c\kappa/n_{eff} = \omega |F| \chi_{fast}^{(3)} A_3^2/2n_{eff}^2$. It is easy to see that for the same pump intensity, the coupling coefficient is significantly less than in the case of the second order nonlinearity.

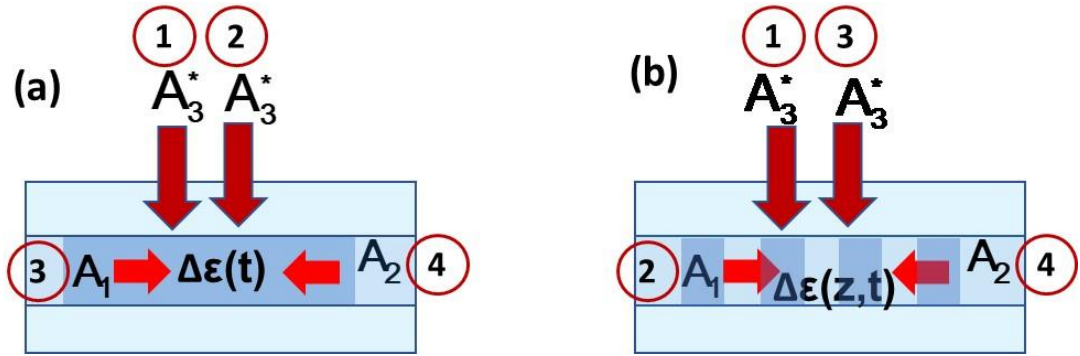

**Figure 6.** Two distinct mechanisms of PTC formation. (a) "fast" mechanism in which two pump photons (1) and (2) modulate the permittivity, and the signal photon (3) scatters into the idler (4). (b) "slow"

mechanism in which one pump (1) and one signal (2) photon produce a moving grating at the beat frequency, and another pump photon (3) scatters off it into the idler (4)

But what about the other two terms (equal to each other) that also engender

$$P_{NL}^{slow} = 2\varepsilon_0 \chi^{(3)}(2\omega_0 - \omega, \omega_0, -\omega, \omega_0) A_3^* e^{i\omega_0 t} A_1 e^{ikz - i\omega t} A_3^* e^{i\omega_0 t} e^{2ik_3 x} \tag{25}$$

This equation describes the process in which first the pump and signal wave mix and produce intensity oscillating at beat frequency $\omega_0 - \omega$, $I_{beat} \sim A_1 A_3^* e^{ikz - (\omega - \omega_0)t}$. If the beat frequency is comparable, or less than the absorption bandwidth, then when the beat wave is absorbed, a slow-moving permittivity grating

$$\Delta\varepsilon_{NL}(z,t) \sim \chi_{slow}^{(3)} A_1 A_3^* e^{ikz - (\omega - \omega_0)t} \tag{26}$$

is formed as shown in Fig.6b. Following that, another pump photon scatters from the moving grating resulting in an idler wave. The mechanism of permittivity change can be associated with absorption saturation via Kramers-Kronig relations[31], or, as is the case of intraband absorption in metals and doped semiconductors, including transparent conductive oxides (TCO), it can be due to increase of the local electron temperature[32-34]. Either way, these processes have characteristic lifetimes over which the changes in permittivity accumulate. For absorption saturation, the characteristic time $\tau$ is the recombination time measured in the range of a few picoseconds[35] to nanoseconds. For intersubband processes in quantum wells this relaxation time is sub-picosecond[36, 37]. In the case of TCO's the time is the temperature relaxation time due to energy transfer from electrons to phonons. This time is measured in hundreds of femtoseconds[38-40]. Naturally, the maximum change of permittivity is proportional to $\tau$, and is thus much larger than the ultrafast change. It has been shown [29, 41] that the relation between fast and slow nonlinearities is that between coherence time $T_2$ and relaxation time $T_1$, which is typically orders of magnitude longer. In case of TCO's the ratio is that between the momentum scattering time $\tau_m$, which is on the scale of a few femtoseconds and the aforementioned electron-lattice relaxation time, measured in hundreds of picoseconds. It is for his reason the third order susceptibility in TCO materials has been measured in the range of $3 \times 10^{-17} m^2 / V^2$ [34, 41]. With that, one can achieve fairly large value of coupling $\kappa_i = \omega | F | \chi_{slow}^{(3)} A_3^2 / 2n_{eff}^2$. However, this assumes that the beat frequency is less than $1/\tau$ - the "real bandwidth" can be found by finding the nonlinear index change from the equation.

$$\frac{d\Delta\varepsilon_{NL}}{dt} = \frac{1}{\tau}\left(-\Delta\varepsilon_{NL} + \chi_{slow}^{(3)} A_p^* A_1 e^{ikz + i\Delta\omega t}\right) \tag{27}$$

This equation has a solution ,

$$\Delta\varepsilon_{NL}(t) = \chi_{slow}^{(3)} A_p^* A_1 \, e^{ikz - i\Delta\omega t} \frac{1 - e^{i\Delta\omega t - t/\tau}}{1 - i\Delta\omega\tau} \tag{28}$$

which indicates the coupling and hence bandgap increasing with time as

$$\kappa_t(t) = \omega \mid F \mid \chi_{slow}^{(3)} A_3^2 / 2n_{eff}^2 \frac{1 - e^{i\Delta\omega t - t/\tau}}{1 - i\Delta\omega\tau} = \kappa_{t,\max}\left(1 - e^{i\Delta\omega t - t/\tau}\right) \tag{29}$$

Therefore, as long as the time constant is comparable or less than modulation time interval T (which in its turn is less than propagation time), one can still see all the salient features of PTC as coupling coefficient increases during the pulse. This is shown in Fig.7 for the case of $\Delta\omega = \kappa_t / 2$ and different values of characteristic times $\tau$. As one can see, the performance deteriorates for longer times. A grating simply does not have enough time to form before the pump shuts down at time T.

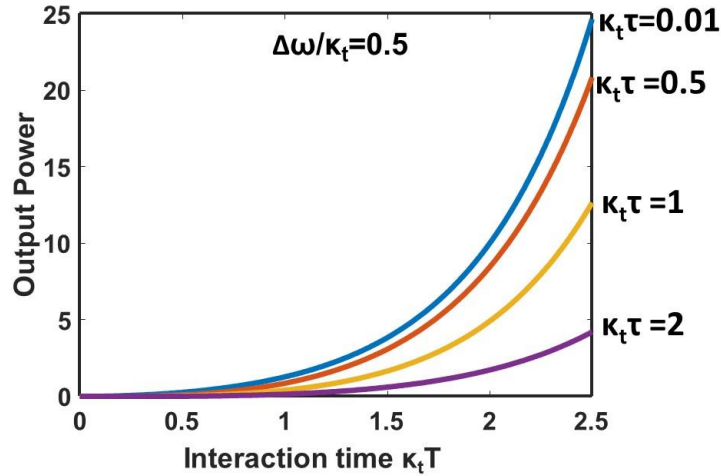

**Figure 7** Performance of PTC-analogue based on "slow" nonlinearity with characteristic response times $\tau$.

It is obvious that the process based on a slow nonlinearity, recently considered in [42] is essentially standard FWM[14] or PC[43] – the only difference being that, instead of two counterpropagating pump waves, one uses co-propagating ones with phase matching achieved due to short propagation distance in the transverse direction. It is instructive to relate the PTC formation using a slow nonlinearity with recording and instant reading of a hologram[44]. Indeed, the signal plays the role of the object wave, while the pump plays the role of the reference wave. Their interference records a dynamic hologram (grating) by modulating the refractive index. This hologram is then instantly read by the pump waves which not only restores the object wave (image) but also produces the conjugate image – idler. Indeed in the experiments on time reversal [45] with TCO's, two waves have been observed, even though the nonlinearity was "slow". It is not surprising to see the gain in this interpretation – obviously using a high brightness restoring wave will increase the brightness of image beyond that of object.

In the end, one can state that ultrafast nonlinearity is not necessarily for observing the phenomena associated with exponential temporal growth inside the bandgap.

## 7. Conclusions

One thing that I definitely did not want to address in this discourse is the question of what all of this portends for practical applications? A lot of enticing prospectives have been envisioned by many authors whose vision extends way further beyond the horizon than my limited eyesight allows me, and a lot of exciting forecasts will be made, no matter what opinion I may express here. So, rather than provoking the righteous ire in the community, I will just summarize a few fundamental facts discovered (or, perhaps, re-discovered) in this work and leave it to the reader to interpret them in practical context. Here are the facts:

1. The PTC is in its heart a parametric process in which modulation of permittivity using second or third order optical nonlinearity causes simultaneous generation of signal and idler photons with energy conservation maintained. The main difference between PTC and conventional parametric processes – OPA, PC, FWM is in the boundary conditions. In conventional parametric processes signal and idler frequencies remain unchanged inside and outside modulated region as only wavevectors can change. For PTC the situation at the temporary boundary is the opposite and while wavevector is maintained before, during, and after time modulation interval, the frequency changes and the conjugated (or time reversed) wave outside the modulation interval have the same frequency as the incident signal. While dispersion curves for OPA and PTC are identical, in PTC one can couple the signal into the bandgap and achieve exponential amplification. That being said, the overall amplification is similar for OPA and PTC – slowly decaying as signal frequency is detuned from the central frequency $\omega_0$. For moderate values of modulation, the OPA actually holds advantage over PTC due to the presence of feedback in space that is obviously impossible in time. It is only when modulation becomes very strong that one can potentially observe the salient feature of PTC – strong amplification within bandgap and almost no amplification outside of it.

2. In terms of amplification bandwidth the PTC in which signal and idler are counterpropagating it will not be easy to match the performance of existing conventional OPAs with copropagating waves in which the parametric gain bandwidth can exceed an octave when dispersion engineering is employed. Counterpropagating scheme does allow wider angular range which may become relevant when amplifying spontaneous emission.

3. In order to observe most of the features of PTC it is not really necessary to operate with ultrafast nonlinearity. Using a relatively slow but strong nonlinearity with response time on the scale of propagation time, i.e. anywhere from a few hundreds of femtoseconds to a few picoseconds (which can be obtained in TCO, a low-temperature growth semiconductor, or intersubband transition in a quantum well) will provide one with a simple way to get all the input-output characteristics of PTC without actually having a bandgap in k-space. Note also, that using this slow nonlinearity and operating near the band edge one can explore an interesting region of the "fast light" occurring near the exceptional point, where the group velocity $d\omega/dk$ approaches very large values and the density of photonics states exhibits singularity[46].

With that, I hope that this essay does help to clear some fundamental uncertainties existing within the community of researchers working in the exciting field of time modulated media. The work performed here may also turn out to be useful in the development of practical time modulated schemes based on second order and slow third order nonlinearities, so it is up to the experimentalists to test and ,with any luck, exploit the conclusions made here.

### Acknowledgements

As always, the contributions of Prof. P. Noir and Dr. S. Artois of JHU, who provided me with all the support and encouragement I needed, are greatly appreciated.

### Conflict of Interest

The author declares no competing financial interest.